\documentclass[twocolumn,showpacs,superscriptaddress,preprintnumbers,amsmath,amssymb,floatfix,prb]{revtex4-1}
\usepackage{graphicx}
\usepackage{dcolumn}
\usepackage[usenames,dvipsnames]{color}
\usepackage{bm}
\usepackage{ulem}
\usepackage{natbib}
\begin{document}
\title{Unusual Structure and Magnetism in MnO Nanoclusters}    
\author{Shreemoyee Ganguly}
\affiliation{Advanced Materials Research Unit and Department of Materials Science, S.N. Bose National Center for Basic Sciences, JD Block, Sector III, Salt Lake City, Kolkata 700 098, India}
\author{Mukul Kabir}
\altaffiliation{Corresponding author: mukulkab@mit.edu}
\affiliation{Department of Materials Science and Engineering, Massachusetts Institute of Technology, Cambridge, Massachusetts 02139, USA}
\author{Biplab Sanyal}
\affiliation{Division of Materials Theory, Department of Physics and Astronomy, Uppsala University, Box 516, SE-75120 Uppsala, Sweden}
\author{Abhijit Mookerjee}
\affiliation{Advanced Materials Research Unit and Department of Materials Science, S.N. Bose National Center for Basic Sciences, JD Block, Sector III, Salt Lake City, Kolkata 700 098, India}
\date{\today}

\begin{abstract}
We report an unusual evolution of structure and magnetism in stoichiometric MnO clusters based on  an extensive and unbiased search through
 the potential energy surface within density functional theory. The smaller clusters, containing up to five MnO units, adopt two-dimensional structures; and regardless of the size of the cluster, magnetic coupling is found to be antiferromagnetic in contrast to previous theoretical findings. Predicted structure and magnetism are strikingly different from the magnetic core of Mn-based molecular magnets, whereas they were previously argued to be similar. Both of these features are explained through the inherent electronic structures of the clusters.  
\end{abstract}
\maketitle

Transition metal oxide, especially MnO~\cite{Nature.365, Nature.416, Nature.409, *Science.2004, *Jacs.2010, Ruettinger.122, Jiao.2010}, clusters have recently attracted extensive multidisciplinary research activity because of their diverse and tunable magnetic and catalytic properties.  Generally, as compared to the bulk, the local magnetic moment is enhanced in  smaller dimensions due to reduction in the number of neighboring atoms. This results in either an overall enhancement of the total moment for the ferromagnetic (FM) case or lead to a finite moment
 even for the  antiferromagnetic (AFM) case due to unequal compensation of spin up and down electrons. Magnetic coupling
 also evolves with particle size, and such size evolution for MnO clusters is non-monotonic.  MnO clusters with a diameter of 5-10 nm show FM behavior~\cite{Li.1997, *Lee.2002} even though the bulk phase is AFM~\cite{PhysRev.76.1256.2, *PhysRevB.50.5041}.
 In contrast, Mn-based single molecular magnets (with magnetic core  $<$  1.5 nm) show a `layered' AFM/ferrimagnetic structure within the mixed-valent Mn centers, resulting in a large magnetic moment and spin anisotropy~\cite{Nature.365, Nature.416}. Moreover, the MnO clusters take essential part in a variety of biological (catalytic) processes from photosynthesis to bacterially mediated organic matter decomposition. The active inorganic center of the oxygen evolving photosystem II contains a manganese oxide cluster (Mn$_4$O$_4$Ca), which catalyzes the light-driven oxidation of water~\cite{Nature.409, *Science.2004, *Jacs.2010}.  Indeed, synthetic complexes containing cuboidal Mn$_4$O$_4$ cores have been found to exhibit unique reactivity in water oxidation/O$_2$ evolution ~\cite{Ruettinger.122}.

The prediction of geometry at the atomic level is one of the most fundamental challenges in condensed matter science. The magnetic and catalytic properties (i.e., broadly speaking: the electronic structure) are strongly coupled to  the `inherent structure'
 (corresponding to minima of the potential energy surface (PES)) of the cluster.  Experimental evidence of structure
 and magnetic coupling, and their size evolution for the transition metal oxide clusters in the gas phase are scarce.
 However, the structure and (FM) magnetism of  (MnO)$_x$ clusters have been  predicted theoretically~\cite{PhysRevLett.81.2970, *Nayak.1999,
 PhysRevB.59.R693, Mithas.2009}. Such a theoretical prediction is complex, and requires a systematic and rigorous search 
through the PES. This is essential  to predict the deepest minima. The  complexity of the PES search increases with 
increasing cluster size. Possible geometrical structures increase exponentially with cluster size, and for a given 
geometrical structure containing $N$ magnetic ions, there are 2$^N$ spin configurations (which may or may not be reduced
 depending on the symmetry). In contrast, all the previous theoretical attempts have been largely biased \cite{PhysRevLett.81.2970, *Nayak.1999, PhysRevB.59.R693, Mithas.2009}, because: (i) geometrical structures were restricted by a particular
 symmetry, and (ii) magnetic structures were restricted to the FM regime. Such limited considerations search only a small
 subspace of the entire PES and thus, previously reported geometric/magnetic structures may not represent the true 
`ground states'. Indeed, in contrast to the previous theoretical reports, in this communication we shall report,
based on a rigorous PES search, that the (MnO)$_x$ clusters show AFM coupling and also show unusual two-dimensional
 (2D) structures up to a certain size.

The spin-polarized density functional theory calculations were conducted using the VASP code~\cite{PhysRevB.47.558, *PhysRevB.54.11169} with the Perdew-Burke-Ernzerhof exchange-correlation functional~\cite{PhysRevLett.77.3865} and the projector 
augmented wave pseudopotential~\cite{PhysRevB.50.17953} at a energy cut-off of 270 eV. Simple cubic supercells were used
 with periodic boundary conditions, and it was made sure that that two neighboring image clusters were separated by at 
least 10 \r A of vacuum space. This ensured that the interaction of the cluster with its periodic image was negligible
 and  reciprocal space integrations were carried out at the $\Gamma$ point.
We started with high symmetry structures, for example, a cubic structure (core of the Mn$_{12}$-molecular magnet~\cite{Nature.365}) for the (MnO)$_4$ cluster. Spin-polarized Born-Oppenheimer molecular dynamics (BOMD) simulations at 1200 K were 
done for 20-30 ps to search for the lowest energy isomers. This approach could efficiently explore the PES.  Several minimum 
energy structures were then picked from these BOMD simulations and were carefully restudied. All of these minimum energy structures were further optimized ({\it local} ionic relaxation) with all possible spin multiplicities. Moreover, we have also considered different spin arrangements for different atoms in the cluster for a particular spin multiplicity.

\begin{figure}[!t]
\begin{center}
\includegraphics[width=7cm, keepaspectratio, angle=0]{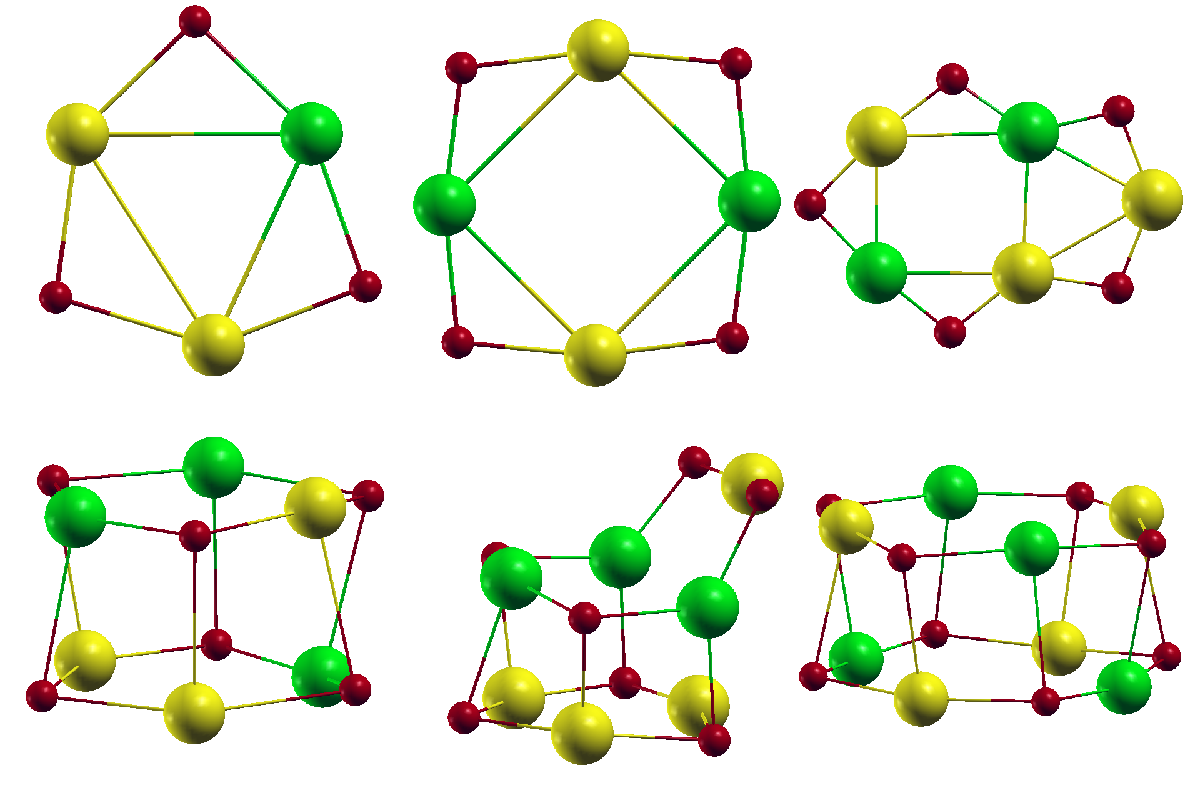}
\caption{\label{structure}(color online) Geometrical evolution shows a 2D to 3D transition at (MnO)$_6$. Minimum energy structures are always found to be antiferromagnetic. The up (down) Mn-atoms are shown with yellow (green) color. Oxygen atoms are shown in red. The Mn-O (1.80 - 2.18 \r A) and Mn-Mn (2.55 - 3.10 \r A) distances, and the Mn-O-Mn angles (78$^{\circ}$-108$^{\circ}$) in these clusters are comparable to the Mn-based molecular magnets~\cite{Nature.365, Nature.416}. }
\end{center}
\end{figure}

\begin{figure}[!t]
\begin{center}
\includegraphics[width=4.2cm, keepaspectratio, angle=270]{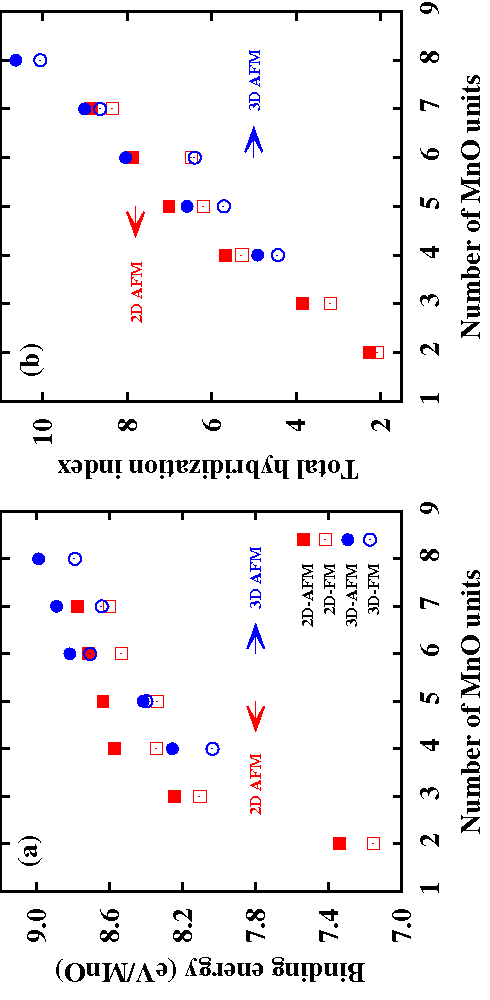}
\caption{\label{binding}(color online) (a) The trend in binding energy shows that the clusters adopt 2D structures until they contain five MnO units, above which they are 3D. The Mn-Mn magnetic coupling is always antiferromagnetic for the most stable solution. Moreover, the AFM coupling is always favorable over the FM coupling, in both 2D and 3D. (b) Total hybridization index,  $\mathcal{H} = \mathcal{H}_{sp} + \mathcal{H}_{pd} + \mathcal{H}_{sd}$, shows a direct correspondence to the energy.}
\end{center}
\end{figure}

Although the magnetic structure is strongly coupled with  cluster geometry, we first discuss the geometric evolution 
alone. For the MnO dimer, as expected,  the Mn-O distance is much smaller (1.65 \r A) than bulk, in agreement with experimental ~\cite{Merer1989} and diffusion Quantum Monte Carlo (DMC) results~\cite{wagner.2007}. Calculated Mn-O stretching frequency (920 cm$^{-1}$) is slightly higher than the experimentally obtained values (832 cm$^{-1}$ Ref.~\cite{Merer1989} and 899 cm$^{-1}$ Ref.~\cite{Hocking.1980}) in the gas phase.   The average Mn-O distances increase with cluster size, and the lowest energy structures are shown in Fig.~\ref{structure}.
Interestingly, the presence of oxygen alters the geometry of MnO cluster as compared to pure Mn clusters~\cite{Kabir.2006}. Moreover, we find that clusters containing up to five MnO units exhibit  two-dimensional (2D) structures (Fig.~\ref{structure}), and 
 the lowest-lying 3D structures are separated from these by a large energy difference [Fig.~\ref{binding}(a)]. 
The structural energy difference $\Delta E_{\rm 2D-3D}$  for (MnO)$_4$ and (MnO)$_5$ clusters are 1.28 and 1.11 eV (equivalent to internal vibrational temperatures of 1650 and 1075 K), respectively. However, the scenario is reversed from the (MnO)$_6$ cluster (Fig.~\ref{structure}), for which $\Delta E_{\rm 3D-2D}$ is 0.61 eV (472 K). The present findings contradict  previous theoretical
 reports, where the PES search were highly biased~\cite{PhysRevLett.81.2970, *Nayak.1999, PhysRevB.59.R693}. In contrast, DMC calculations {\it biased with only FM coupling} predicts a similar geometrical trend for $x \le 4$~\cite{Mithas.2009}. 
The present results are in accordance with the prediction of mass spectra~\cite{Zeimann.1992}, where Ziemann and Castleman 
proposed non-cubic structural growth, and larger clusters were composed of relatively more stable hexagonal (MnO)$_3$ and
 rhombic (MnO)$_2$ units. Similarly, we also find that the (MnO)$_3$ cluster is more stable (magic cluster) and serves as 
the building block for larger clusters. It was believed earlier that the small MnO clusters serve as the magnetic core
 of Mn-based single molecular magnets ~\cite{PhysRevB.59.R693}; the present study confirms that there is no such structural
 resemblance in terms of symmetry. The planar structure of (MnO)$_4$ (Fig.~\ref{structure}) is substantially different 
from the magnetic core of [Mn$_4$O$_3$Cl$_4$(O$_2$CEt)$_3$(py)$_3$]$_2$~\cite{Nature.365}, [Mn$_{12}$O$_{12}$(CH$_3$COO)$_{16}$(H$_2$O)$_4$]~\cite{Nature.416} molecular magnets and the oxygen evolving center (Mn$_4$O$_4$Ca) in photosystem II
 ~\cite{Nature.409}. The structural symmetry of (MnO)$_6$ is very different from that observed in octahedral 
[Mn$_6$($\mu_6$-O)($\mu$-OR)$_{12}$] molecular complex~\cite{Stamatatos20091624}. However, the structural parameters 
such as Mn-O (1.80-2.18 \r A), Mn-Mn (2.55-3.10 \r A) bond lengths, and Mn-O-Mn angles (78$^{\circ}$-108$^{\circ}$) 
in these clusters are similar as compared to the molecular magnets (1.85-2.23 \r A, 2.77-3.44 \r A, and 
94$^{\circ}$-137$^{\circ}$, respectively)~\cite{Nature.365, Nature.416}. 
It should also be noted that the Mn-O distance in these clusters already approaches that of the bulk MnO (2.25 \r A).

The binding increases substantially due to the presence of oxygen (Fig.~\ref{binding}). This can be explained in terms of
 electronic configurations. The free Mn atom has 3$d^5$4$s^2$ electronic configuration, and thus the Mn-dimer is weakly 
bonded~\cite{Kabir.2006}. In contrast, our Bader charge analysis~\cite{Bader, *Henkelman2006354} shows that Mn atoms lose 
1.20$e$ charge to O in  (MnO)$_2$, which makes binding in these clusters stronger. This amount ($\sim$1.20$e$) of charge
 transfer remains the rule of thumb for all the minimum energy structures for the stoichiometric MnO clusters. 
In addition to the covalency, this significant charge transfer suggests an ionic contribution to the Mn-O bonding.

Calculated stability, as measured by $\Delta E_x = E({\rm MnO})_{x-1} + E({\rm MnO})_{x+1} - 2E({\rm MnO})_x$, where $E$ is
 the total binding energy, shows local peaks for clusters with $x=$ 3, 4 and 6, in agreement with 
experiment~\cite{Zeimann.1992}. This indicate exceptional stability referring to their magic nature. Clusters with three and four MnO units serve as the building blocks for larger clusters. Two (MnO)$_3$ cluster units are stacked in 3D for the ground state (Fig.~\ref{structure}), when the similar stacking in 2D (with similar AFM magnetic ordering) is 0.61 eV higher in 
energy [Fig.~\ref{binding}(a)]. Similarly, as shown in Fig.~\ref{structure}, a (MnO)$_3$ unit is stacked in 3D with
 a distorted (MnO)$_4$ unit to form the most stable (MnO)$_7$ cluster. Finally, the (MnO)$_8$ cluster is formed by staking  two (MnO)$_4$ units.

\begin{figure}[!t]
\begin{center}
\includegraphics[width=8.5cm, keepaspectratio, angle=0]{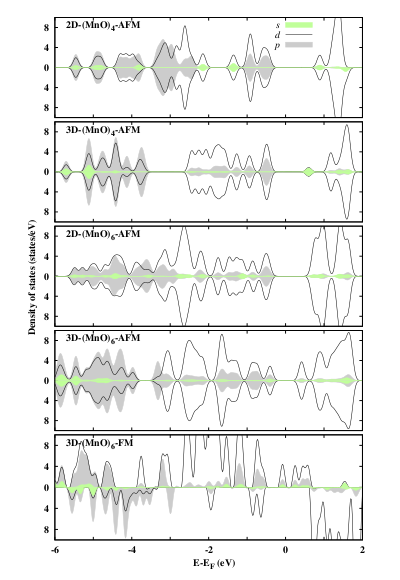}
\caption{\label{DOS}(color online) Orbital projected density states summed over all the atoms in the cluster (for 2D and 3D (MnO)$_4$ and (MnO)$_6$ clusters in their AFM state; FM state for 3D (MnO)$_6$ is also shown) show that the minimum energy structures (both geometric and magnetic) correspond to larger energy spread of the orbitals and higher hybridization.  Gaussian smearing (0.1 eV) has been used. }
\end{center}
\end{figure}

To understand this morphological transition, we plot the orbital projected density of states (DOS) for the clusters 
(Fig. \ref{DOS}). These show a clear trend: the energy spread of the orbitals is higher for the minimum energy structure,
 and also has a larger orbital overlap.  For example, a (MnO)$_4$ cluster has larger orbital spread and also has higher $s$-$d$ and $p$-$d$ hybridization in 2D compared to the respective optimal 3D structure. In contrast, the situation is reversed
 for (MnO)$_6$ cluster, where the larger spread and higher $s$-$d$ and $p$-$d$ hybridization makes the 3D structure 
 favorable over the 2D one. The orbital hybridization can be quantified, and could explain the cluster morphology.~\cite{Hakkinen.2002} We calculate the ($k$-$l$) hybridization index,
\begin{equation}
\mathcal{H}_{kl} = \sum_I \sum_i w_{ik}^I w_{il}^I,
\label{equation}
\end{equation}
where $k$, $l$ are the orbital indices,  $w_{ik}^I (w_{il}^I)$ is the square projection of the $i$-th Kohn-Sham orbital on
 to the $k$ ($l$) spherical harmonics centered at atom $I$ and integrated over an atomic sphere (radius depends on the atom
 type, Mn/O~\cite{Note1}). Note that the spin index is inherent. However, unlike gold clusters, in a system with active
 O-$p$ electrons,~\cite{Hakkinen.2002} in addition to the $s$-$d$ hybridization index ($\mathcal{H}_{sd}$), the $\mathcal{H}_{pd}$ and $\mathcal{H}_{sp}$ would also play an important role for determining the dimensionality. Indeed, we find [Fig. \ref{binding}(b)] that the total hybridization index  $\mathcal{H}$ is always higher for the lowest energy structures for the entire size range.

Next we turn our attention to the magnetic ordering. We find that the Mn-atoms in all clusters are AFM coupled. This gives rise to a net 0 or 5 $\mu_B$ moment for the clusters with even and odd number
 of MnO units, respectively. It is also interesting to note here that regardless of their dimensional nature, the AFM 
arrangement is always energetically favorable [Fig.~\ref{binding}(a)].  Experimental evidences of magnetic 
structure in such clusters are scarce, but the calculated results for the MnO dimer (sextet) is in agreement with the 
only available experiment ~\cite{Merer1989}. Although the moment is localized on the Mn-sites, a small $p$ polarization (0.02-0.15 $\mu_B$) is observed for the oxygen atoms. These results of the magnetic structure are in
 direct contradiction with the previous results including the DMC calculations~\cite{PhysRevLett.81.2970, *Nayak.1999,
 PhysRevB.59.R693, Mithas.2009}. As mentioned before, all the previous studies were highly biased and only scanned a small
 subspace of the PES constrained either by geometrical structure with high symmetry or by FM coupling, or constrained 
by both.  We find that the presence of oxygen stabilizes  Mn-Mn AFM coupling. For example, the energy difference $\Delta E({\rm FM-AFM})$ is only -0.2 and 0.14 eV for pure Mn$_4$ and Mn$_6$ clusters respectively~\cite{Kabir.2006}, whereas this difference substantially increases to 0.92 and 0.67 eV, respectively for the (MnO)$_4$ and (MnO)$_6$ clusters. It would be interesting to compare the magnetic structure of 
these clusters with the [Mn$_{12}$O$_{12}$(CH$_3$COO)$_{16}$(H$_2$O)$_4$] molecular magnet~\cite{Nature.416}. 
The Mn atoms in the inner shell (four Mn$^{4+}$) and in the outer shell (eight Mn$^{3+}$) are FM coupled within both 
of the shells, but are AFM coupled between the shells. This is strikingly different from the present (MnO)$_x$ clusters in the gas phase.

Similar to the morphological stability, the magnetic stability  can also be explained in terms of the total hybridization
 index [Fig.~\ref{binding} (b)] and projected density of states (Fig.~\ref{DOS}). Compared to the FM structure, the AFM structure has larger orbital spread,
 and also accompanied by higher hybridization index. For example, the 3D-FM structure for (MnO)$_6$ has sharp $d$-states 
(Fig.~\ref{DOS}), and in contrast, the $d$-states for 3D-AFM (MnO)$_6$ are more widely spread in energy and show larger overlap with O-$p$ states. Calculated total hybridization index $\mathcal{H}$, as shown in Fig.~\ref{binding}(b), is much higher in the AFM state (8.04) than in the FM state (6.40) in 3D, confirming the preference of AFM coupling.  We restudied the (MnO)$_4$ and (MnO)$_6$ clusters using hybrid PBE0 functional for the exchange-correlation, and we should point out that both the structural and magnetic trends discussed here are found to be unaltered.

Compared to AFM bulk MnO crystal~\cite{PhysRev.76.1256.2, *PhysRevB.50.5041}, the overall exchange mechanism is 
complicated in MnO clusters. The semi-empirical Goodenough-Kanamori rules can be employed to understand the Mn-Mn 
magnetic coupling. In addition to the Mn-O-Mn superexchange mechanism, direct Mn-Mn exchange mechanism is also present
 in these clusters since the Mn-O-Mn angle is much smaller than 180$^{\circ}$ (Fig.~\ref{structure}). As we have 
discussed earlier,  pure Mn$_4$ is FM with $\Delta E({\rm FM-AFM})$ = -0.2 eV, i.e., direct exchange prefers FM coupling.
 When O atoms are introduced, the magnetic structure changes to AFM due to stronger 
superexchange coupling that prefers AFM Mn-Mn ordering. In contrast, the direct exchange in pure Mn$_6$ is already AFM, and is further stabilized due to the 
AFM superexchange when O is introduced. Similar to the ferrimagnetic Fe$_4$O$_6$ clusters~\cite{PhysRevB.82.020405}, 
these stoichiometric MnO clusters also have a very large magnetic exchange (at least 0.11 eV per MnO unit), 
which is much larger than the Mn-based molecular magnets~\cite{Nature.365, Nature.416}. Therefore, the Curie temperature of these (MnO)$_x$ clusters would be much higher than the corresponding AFM bulk MnO ($\sim$ 118 K). This can be exploited to tailor new materials.

In summary, we have demonstrated, through a rigorous and unbiased potential energy search, that the smaller stoichiometric MnO 
clusters show unusual 2D structures, and that Mn atoms are AFM coupled.  Both these features are explained in terms 
of the inherent electronic structure of these clusters. Present results deviate from the earlier theoretical 
predictions as those works  explored only a small subspace of the potential energy surface constrained by the high 
symmetry structures and ferromagnetic  coupling~\cite{PhysRevLett.81.2970, *Nayak.1999, PhysRevB.59.R693, Mithas.2009}. 
Although the experimental results on such clusters are scarce, the present results agree well with the limited experimental predictions on the cluster structure and stability~\cite{Zeimann.1992}. However, there is no experimental evidence on 
the evolution of magnetic structure; we believe the complementary infrared dissociation spectroscopy~\cite{PhysRevB.82.020405} will be helpful to confirm  both the geometric and magnetic structures of such transition metal oxide
 clusters in the gas phase.

M. K.  gratefully acknowledges helpful discussions with D. Ceresoli, L.-L. Wang, K. Doll, and D. G. Kanhere. B. S. acknowledges financial support from Swedish Research Links program funded by VR/SIDA.

\end{document}